\documentclass[twocolumn,pre,showpacs]{revtex4-1}
\usepackage{graphicx}
\usepackage{amsmath}
\usepackage{amsfonts}
\usepackage{mathtools}
\usepackage{subfigure}
\usepackage[utf8]{inputenc}

\begin{document}
\title{Skewness and Kurtosis in Statistical Kinetics}

\author{Andre C. Barato$^{1,2}$ and Udo Seifert$^{1}$}
\affiliation{ $^1$ II. Institut f\"ur Theoretische Physik, Universit\"at Stuttgart, 70550 Stuttgart, Germany\\
             $^2$ Max Planck Institute for the Physics of Complex Systems, N\"othnizer Straße 38, 01187 Dresden,Germany}

\parskip 1mm
\def\F{\mathcal{A}} 
\def\Q{\mathcal{Q}} 
\def\R{{\mathcal{R}}}
\def\S{{\mathcal{S}}}
\def\K{{\mathcal{K}}}
\def\D{{\mathcal{D}}}
\def\i{\text{\scriptsize $\cal{I}$}}

\begin{abstract}
We obtain lower and upper bounds on the skewness and kurtosis associated with the cycle completion time of 
unicyclic enzymatic reaction schemes. Analogous to a well known lower bound on the randomness parameter, the lower bounds 
on skewness and kurtosis are related to the number of intermediate states in the underlying chemical reaction
network. Our results demonstrate that evaluating these higher order moments with single molecule data can lead 
to information about the enzymatic scheme that is not contained in the randomness parameter.
\end{abstract}


\maketitle


In enzyme kinetics one typically studies the relation between the average rate of product 
formation and parameters like the substrate concentration \cite{corn13}. If the number of enzymes 
is large, fluctuations of this rate of product formation are small and can be neglected. 
Many different enzymatic schemes can lead to the same phenomenological expression for the rate 
of product formation as a function of the substrate concentration. Therefore, in general, 
this quantity is not sufficient to provide essential information about the underlying enzymatic scheme, 
like, for example, the number of intermediate states.

Statistical kinetics \cite{moff14,moff10,schn95,shae05,li13} is a novel field where one tries to infer the underlying chemical
reaction scheme from measurements of the, typically, large fluctuations in single molecule experiments \cite{svob93,rito06,gree07,corn07,moff08,xie13}.
For example, experimental data on the fluctuations related to the time to complete an enzymatic cycle have been used to 
infer properties of the enzymatic mechanism  of the packing motor in bactheriophage $\phi$29 \cite{moff09}. Related analysis for
kinesin and myosin can be found in \cite{schn97,viss99,valen06,fish01,kolo03}.

The time to complete an enzymatic cycle is a key quantity in statistical kinetics.
The size of its fluctuations is quantified by the randomness parameter. The inverse 
of this randomness parameter bounds the number of intermediate states of a unicyclic enzymatic scheme \cite{moff14}. 
This main result of statistical kinetics implies that measurements of the randomness parameter lead to 
a lower bound on the number of intermediate states. Further important information that 
can be inferred from this randomness parameter is whether the chemical reaction network 
constitutes of a cycle or is more complex. Specifically, a randomness parameter larger than 1 implies that 
the reaction scheme is not simply a cycle \cite{engl06,kou05,moff10a,saha12,foot}. Furthermore, similar to 
the Michaelis-Menten expression for the rate of product formation, the randomness parameter can also 
be written as a function of the substrate concentration \cite{moff10} (see also \cite{chem08,chau13,jung10}).

The randomness parameter is related to the second moment of the cycle completion time
distribution, i.e., it is the variance divided by the mean square of this distribution. A natural 
and important question for statistical kinetics is whether higher order moments can 
provide further information about the enzymatic scheme. This view is expressed in the recent review by
Moffit and Bustamante \cite{moff14}: ``{\sl... some of the most exciting advances in statistical kinetics 
may come from generalizations of the randomness parameter to higher moments ...}''.

In this letter, we analyze the third and fourth standardized moments associated with the
probability distribution of the cycle completion time, namely, the skewness and kurtosis, respectively. 
We show that similar to the randomness parameter these higher order moments have upper 
and lower bounds in unicyclic enzymatic schemes, with the lower bound depending on the 
number of intermediate states. Skewness and kurtosis can be used to derive a 
stronger lower bound on the number of intermediate states and to identify the presence 
of extra cycles or an extra state in the enzymatic scheme more effectively.


We consider a generic unicyclic enzymatic scheme of the form
\begin{equation}
 1\xrightleftharpoons[k_1^-]{k_1^+} 2 \xrightleftharpoons[k_2^-]{k_2^+} 3\ldots N-1\xrightleftharpoons[k_{N-1}^-]{k_{N-1}^+}N \xrightharpoonup{k_N^+}1.
\label{fullreaction} 
\end{equation}
where $k_i^{\pm}$ are transition rates and the numbers $i=1,2,\ldots,N$ represent different states of an enzyme E. For example, the Michaelis-Menten scheme with 
$N=2$ reads $\textrm{E}+\textrm{S}\xrightleftharpoons[k_1^-]{k_1^+} \textrm{ES} \xrightharpoonup{k_2^+} \textrm{E}+\textrm{P}$.
The last transition is taken to be irreversible, i.e., $k_N^-=0$, and for the reminder of the paper we fix the time-scale by setting $k_N^+=1$. The time evolution of the probability of being
in state $i$ at time $\tau$, $P_i(\tau)$, is governed by the master equation
$d\mathbf{P}/d\tau= \mathbf{L}\mathbf{P}$, where $\mathbf{P}$ is a vector with dimension $N$ and components $P_i(\tau)$. The stochastic matrix $\mathbf{L}$ is given by
the $z=0$ case of the matrix  
\begin{equation}	
[\mathbf{L}(z)]_{ji}\equiv\left\{\begin{array}{ll} 
 k_i^+\textrm{e}^{z\delta_{i,N}}  & \quad \textrm{if } j=i+1\\
 k_i^- & \quad \textrm{if } j=i-1\\
 -k_i^+-k_{i-1}^- & \quad \textrm{if } i=j \\
 0 & \quad \textrm{otherwise,} 
\end{array}\right.\,
\label{modgenerator}
\end{equation}
where $z$ is a real variable, $\delta_{i,N}$ denotes the Kronecker delta, and we assume periodic boundary conditions, i.e.,  
$i+1=1$ for $i=N$ and $i-1=N$ for $i=1$. This $z$-dependent matrix is needed for the calculations below. 

The probability that a cycle is completed at time $t$ is denoted by $\psi(t)$. Its moment of order $n$ is defined as  
\begin{equation}
t_n\equiv \int_0^\infty \psi(t)t^n.
\label{momentdef}
\end{equation}
The main quantity in statistical kinetics is the randomness parameter 
\begin{equation}
\R\equiv \sigma^2/t_1^2,
\end{equation}
where $\sigma\equiv \sqrt{t_2-t_1^2}$. The third and fourth order standardized moments are the skewness 
\begin{equation}
\S\equiv (t_3-3t_1 \sigma^2 -t_1^3)/\sigma^3,
\label{sdef}
\end{equation}
and the kurtosis
\begin{equation}
\K\equiv (t_4-4t_3t_1+6\sigma^2t_1^2+3t_1^4)/\sigma^4.
\label{kdef}
\end{equation}

The moments of $\psi(t)$ can be obtained from the generating function 
\begin{equation}
\tilde{\psi}(s)\equiv\int_0^{\infty}\psi(t)\textrm{e}^{-st}dt
\end{equation}
as 
\begin{equation}
t_n= (-1)^n\frac{d^n\tilde{\psi}}{ds^n}.
\label{momentgen}
\end{equation}
As shown in \cite{chem08} this generating function for the scheme in eq (\ref{fullreaction}) can be calculated explicitly.
The expression for $\tilde{\psi}(s)$ involves the determinant of the matrix $s\mathbf{I}-\mathbf{L}(z)$, where $\mathbf{I}$ is the identity matrix. 
Specifically, in terms of the coefficients $\{c_j(z)\}$, defined through  
\begin{equation}
\textrm{det}\{s\mathbf{I}-\mathbf{L}(z)\}\equiv \sum_{j=0}^N c_j(z) s^j,
\label{coeff}
\end{equation}
the generating function is given by \cite{chem08}
\begin{equation}
\tilde{\psi}(s)= \Gamma_+\left/\left(\sum_{j=1}^N c_js^j +\Gamma_+\right)\right.,
\label{generatingfunc}
\end{equation}
where $\Gamma_+\equiv k_1^+k_2^+\ldots k_N^+$ and $c_j$ for $j\neq 0$ is independent of $z$. This expression comes from a relation between $\psi(t)$ and the
probability of completing $M$ cycles up to time $t$ that is explained in \cite{supp}. The quite involved expression of these coefficients for arbitrary $N$ is provided in \cite{chem08}.
These coefficients can also be calculated from an iterative procedure developed in \cite{li13}.
Hence, we can obtain an expression for the skewness and kurtosis in terms of the transition rates with the following procedure: we calculate the coefficients in eq (\ref{coeff}), 
obtain the generating function with eq (\ref{generatingfunc}) and calculate the first four moments defined in eq (\ref{momentdef}), which finally lead to the skewness and kurtosis with eqs
(\ref{sdef}) and (\ref{kdef}), respectively.

The randomness parameter for the model in eq (\ref{fullreaction}) is bounded by \cite{moff14}
\begin{equation}
1/N\le \R\le 1.
\label{boundrand}
\end{equation}
These bounds are of central importance in statistical kinetics. First, measurements of $\R$ thus lead to 
the lower bound $N_{\textrm{min}}^{\R}\equiv 1/\R$ on the number of intermediate states in the enzymatic scheme. Second, a randomness parameter that is larger than 1 indicates 
that the underlying enzymatic scheme is not just a simple cycle as in eq (\ref{fullreaction}).

As our first main result, we can show that skewness and kurtosis for the generic unicyclic network in eq (\ref{fullreaction}) are also 
bounded. These bounds are given by 
\begin{equation}
2/\sqrt{N}\le \S\le 2
\label{boundskew}
\end{equation}
for the skewness, and 
\begin{equation}
6/N+3\le \K\le 9
\label{boundkurt}
\end{equation}
for the kurtosis. These results are based on strong numerical evidence obtained in the following way. We calculated the skewness and kurtosis 
following the procedure explained after eq (\ref{generatingfunc}) as functions of the $2N-2$ free transition rates from $N=2$ up to $N=8$. With numerical minimization (maximization), 
we obtain that they are minimized (maximized) exactly at the value quoted as lower (upper) bounds in eqs (\ref{boundskew}) and (\ref{boundkurt}). Furthermore,
we have also evaluated $\S$ and $\K$ at randomly chosen transition rates and observe that they fulfill the above bounds. This numerical evidence for the bounds in eqs (\ref{boundskew}) 
and (\ref{boundkurt}) is further discussed in \cite{supp}, where we also provide analytical evidence that includes a proof of the bounds for the case $N=2$.

For all three quantities, randomness parameter, skewness, and kurtosis, the lower bound is reached when all forward rates are $k_i^+=1$ and all backward rates are $k_i^-=0$.
In this case $\psi(t)=t^{N-1}\textrm{e}^{-t}/(N-1)!$ is the Erlang distribution \cite{aldo87}. The upper bounds are reached when the rates are chosen such that $\psi(t)$ becomes the exponential distribution, 
which is the distribution for the case $N=1$.  
 
We now show that evaluating skewness $\S$ and kurtosis $\K$ can lead to further information on the minimal
number of states that can not be obtained from the randomness parameter $\R$. A first presumption might be that the bounds 
$N_{\textrm{min}}^{\S}\equiv 4/\S^2$ or $N_{\textrm{min}}^{\K}\equiv 6/(\K-3)$, as obtained from 
eq (\ref{boundskew}) and eq (\ref{boundkurt}), respectively, could be sharper than  $N_{\textrm{min}}^{\R}$.
However, a thorough numerical investigation up to $N=8$ shows that this presumption is incorrect, and the inequalities     
\begin{equation}
N_{\textrm{min}}^{\R}\ge N_{\textrm{min}}^{\S}\ge N_{\textrm{min}}^{\K},
\end{equation}
hold. Hence, the randomness parameter provides a tighter bound on the number of intermediate states. Moreover, since a precise determination of higher order moments requires better
statistics, using the lower bounds in eqs (\ref{boundskew}) and (\ref{boundkurt}) to estimate the number of states is pointless. 

\begin{table}
\caption{Maximum of differences  for $N=2,3,\ldots,8$.}
\label{tab1}
\begin{center}
\begin{tabular}{l|c|c|c|c|c|c|r}
$N$  		& 2	& 3  & 4 & 5 & 6 & 7 & 8  \\
\hline
$\textrm{max}\{\D_1\}$  & $0.325$	 & $0.692$ & $1.090$ & $1.510$ & $1.950$ & $2.403$ & $2.869$   \\
$\textrm{max}\{\D_2\}$  & $0.073$	 & $0.161$ & $0.261$ & $0.370$ & $0.487$ & $0.610$ & $0.739$   \\
$\textrm{max}\{\D_3\}$  & $0.382$	 & $0.816$ & $1.286$ & $1.782$ & $2.299$ & $2.833$ & $3.382$   
\end{tabular}
\end{center}
\end{table}

\begin{figure}
\includegraphics[width=65mm]{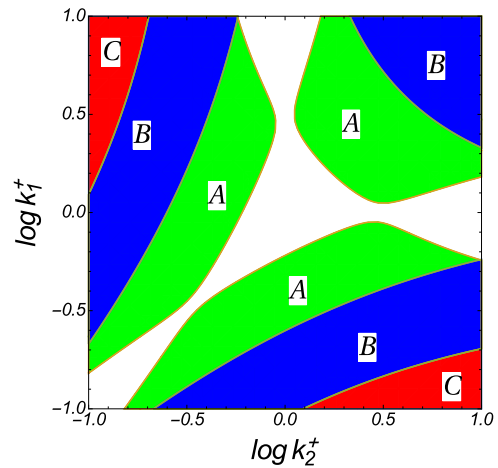}
\vspace{-2mm}
\caption{(Color online) Illustration of the new information provided by $\D_1$. The parameters 
are $N=3$ and $k_1^-=k_2^-=0$. In the blue region marked by $B$, $N_{\textrm{min}}^{\R}<2$ and $\D_1>0.3246$, where $0.3246$
is the maximum value of $\D_1$ for $N=2$; in the red region marked by $C$, $N_{\textrm{min}}^{\R}<2$ and $\D_1<0.3246$; in the green region marked by $A$, 
$N_{\textrm{min}}^{\R}>2$ and $\D_1<0.3246$. The region marked by $B$ corresponds to the manifold in parameter space where evaluating 
$\D_1$ leads to the information that $N$ must be larger than $2$, while the randomness parameter gives a lower bound below $2$. 
}
\label{fig1} 
\end{figure}

Nevertheless, skewness and kurtosis can still be used to obtain further information on the minimal number of states by evaluating the positive differences 
\begin{equation}
\D_1\equiv N_{\textrm{min}}^{\R}-N_{\textrm{min}}^{\S}\ge 0,
\end{equation}
\begin{equation}
\D_2= N_{\textrm{min}}^{\S}-N_{\textrm{min}}^{\K}\ge 0,
\end{equation}
and
\begin{equation}
\D_3=N_{\textrm{min}}^{\R}-N_{\textrm{min}}^{\K}\ge 0.
\end{equation}
Numerical investigation of these differences shows that they have a maximum that depends on the number of states $N$. The values of this maximum for different $N$, which are obtained with numerical maximization, 
are reported in Table \ref{tab1}, which constitutes our second main result. Note that these maxima increase with $N$. Therefore, the upper bounds on the differences $\D$ can be used to bound the number of states $N$ from below. 
For example, if $\D_1>0.3246$, then the number of states must be larger than $N=2$. As shown in Figure \ref{fig1}, for $N=3$ there are regions
in the space of parameters where $N_{\textrm{min}}^{\R}<2$, i.e., the randomness parameter indicates a number of states that is not larger than $2$, 
while the difference $\D_1>0.3246$ indicates $N>2$. Hence, the difference $\D_1$ leads to a stronger lower bound on the number of intermediate states in these regions.   

A posteriori, we understand why these upper bounds on the differences $\D$ can lead to better lower bounds on the number of intermediate states.
The quantities  $N_{\textrm{min}}^{\R}$, $N_{\textrm{min}}^{\S}$, $N_{\textrm{min}}^{\K}$ are equal to $N$ when all forward transition rates
are equal and all backward transition rates are 0. If the transition rates are such that $\psi(t)$ is exponential, then these quantities reach
their minimal value 1. For the differences $\D$ the situation is different. For both of these extreme cases, irreversible transitions with uniform rates and $\psi(t)$
exponential, all differences $\D$ are 0. Therefore, the differences are maximized somewhere in the space of transition rates  where  
$N_{\textrm{min}}^{\R}$, $N_{\textrm{min}}^{\S}$, $N_{\textrm{min}}^{\K}$ are between their extreme values. A precise analytical identification of 
the regions in this space where the differences $\D$ lead to a sharper lower bound on number of states for arbitrary $N$, however, does not seem to be simple.

\begin{figure}
\subfigure[]{\includegraphics[width=65mm]{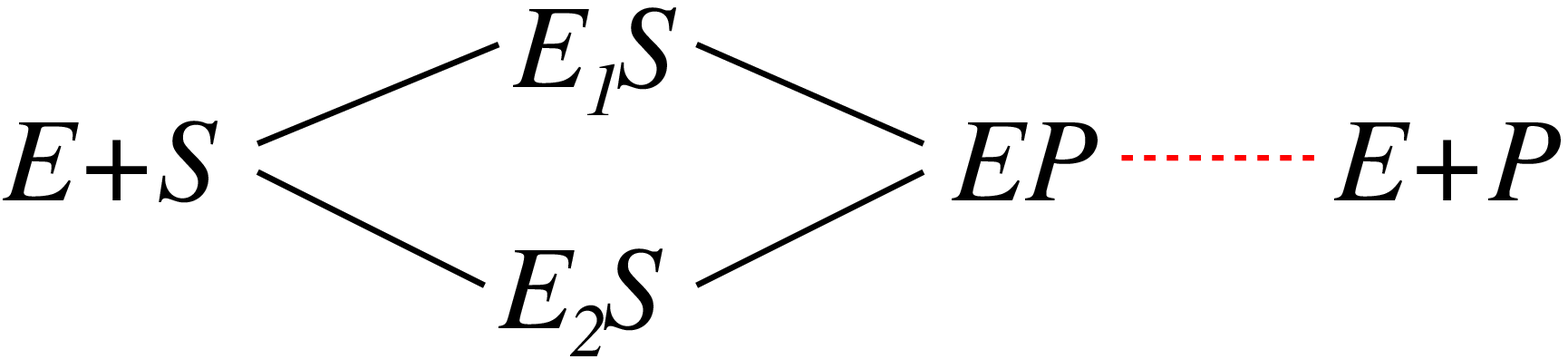}\label{fig1.5a}}
\subfigure[]{\includegraphics[width=65mm]{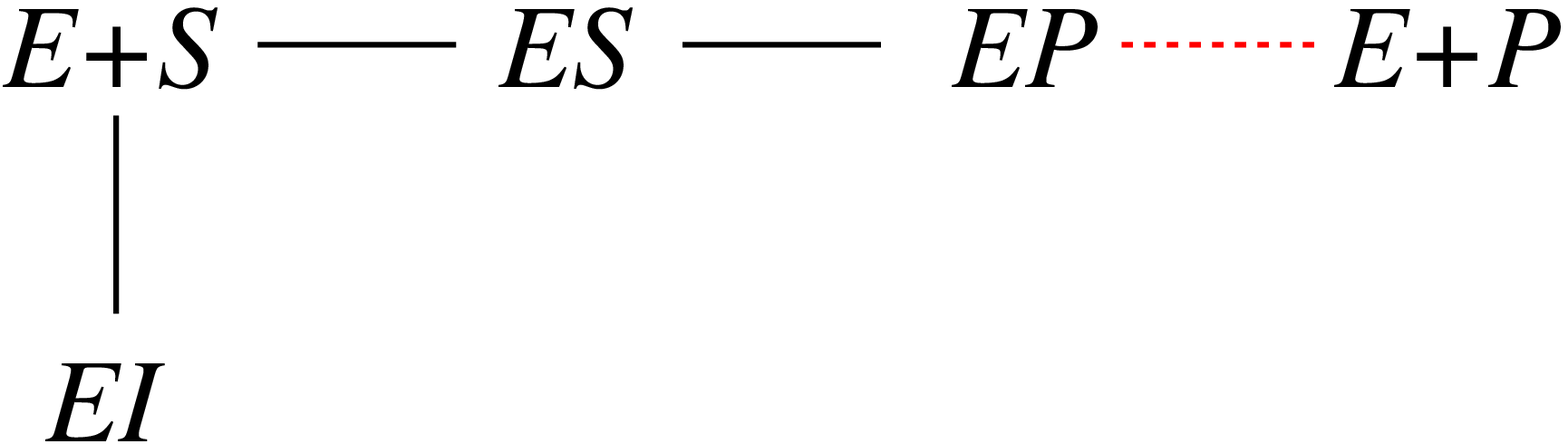}\label{fig1.5b}}
\vspace{-2mm}
\caption{(Color online) Chemical reaction schemes. Scheme (a) corresponds to the multicyclic model in eqs (\ref{reaction1}) and (\ref{reaction2}), and scheme (b) corresponds to the model with an an off-pathway state that is obtained 
by setting the rates $k_{34}=k_{43}=0$. For both schemes, the red dotted link represents the transition that is irreversible.
}
\label{fig1.5} 
\end{figure}

Skewness and kurtosis can also be used to identify more efficiently than the randomness parameter whether the topology of the enzymatic 
network is more complex than a single cycle. As a case study to demonstrate this fact we consider the multicyclic enzymatic network from Figure \ref{fig1.5a}.
There are two intermediate states of the enzyme bound with substrate, which are denoted $\textrm{E}_1\textrm{S}$ and $\textrm{E}_2\textrm{S}$. 
 The generation of product can then happen in two different pathways:
\begin{equation}
 \textrm{E}+\textrm{S}\xrightleftharpoons[k_{21}]{k_{12}}\textrm{E}_1\textrm{S} \xrightleftharpoons[k_{42}]{k_{24}} \textrm{EP} \xrightharpoonup{k_{41}} \textrm{E}+\textrm{P},
\label{reaction1} 
\end{equation} 
and 
\begin{equation}
 \textrm{E}+\textrm{S}\xrightleftharpoons[k_{31}]{k_{13}}\textrm{E}_2\textrm{S} \xrightleftharpoons[k_{43}]{k_{34}} \textrm{EP} \xrightharpoonup{k_{41}} \textrm{E}+\textrm{P},
\label{reaction2} 
\end{equation}
where we identify the states $\textrm{E}\mathrel{\hat=} 1$, $\textrm{E}_1\textrm{S}\mathrel{\hat=} 2$, $\textrm{E}_2\textrm{S}\mathrel{\hat=} 3$, and $\textrm{EP}\mathrel{\hat=} 4$. The method we use to calculate $\R$, $\S$ and $\K$, which have quite
long expressions, for this model is explained in \cite{supp}. 

\begin{figure}
\includegraphics[width=65mm]{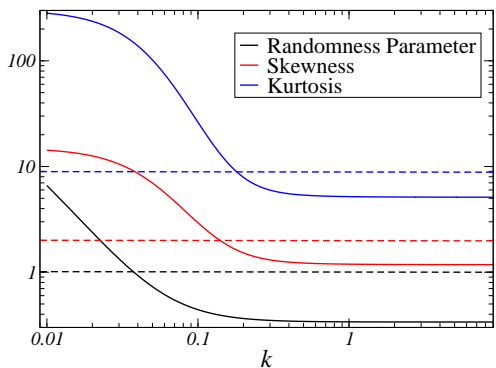}
\vspace{-2mm}
\caption{(Color online) Randomness parameter, skewness, and kurtosis for the model shown in Figure \ref{fig1.5a}. The transition rates 
are $k_{12}=k_{24}=k_{41}=1$, $k_{21}=k_{42}=k_{14}=0$, $k_{13}=k_{43}=10^{-2}$, and $k_{31}=k_{34}=k$. The horizontal dotted lines indicate the
respective maximum values for a unicyclic  network.    
}
\label{fig2} 
\end{figure}

The maxima of randomness parameter $\R$, skewness $\S$ and kurtosis $\K$ for a unicyclic network are $\R=1$, $\S=2$, and $\K=9$, see eqs (\ref{boundrand}), (\ref{boundskew}), and (\ref{boundkurt}).
If any of these quantities exceeds these maximal values 
the enzymatic scheme must be different from a single cycle. In order to check which of these quantities is more effective in providing a signature of a non-unicyclic network we set the rates 
as  $k_{12}=k_{24}=k_{41}=1$, $k_{21}=k_{42}=0$, $k_{13}=k_{43}=10^{-2}$, and $k_{31}=k_{34}=k$. For $k\gg 1$, the cycle in eq (\ref{reaction1}) dominates the network: in the rare occasions a jump to
the state $\textrm{E}_2\textrm{S}$ occurs the enzyme quickly returns to the main cycle. In this case $\R$, $\S$, $\K$ are close to the values they reach in the lower bounds with $N=3$.
If $k\ll 1$, the system can get trapped in state $\textrm{E}_2\textrm{S}$. Such a trapping state makes very large cycle completion times more probable, which leads to an increase 
in fluctuations. As shown in Fig. \ref{fig2}, all three quantities cross their maximum allowed values in a unicyclic network as $k$ gets smaller. The main
result is that the kurtosis is the first to indicate a more complex network followed by the skewness and the randomness parameter. For example, at $k=0.376$ the randomness parameter is
$\R= 1.00$ and the kurtosis is $\K= 148.7$, way above the value $\K=9$. With a numerical investigation in the whole parameter space we find that higher order moments are always more 
effective in revealing whether the network contains a structure more complex than a cycle, which is our third main result. This numerical investigation consists of, for example, maximizing the skewness with the constraint 
that $\R\le 1$, which leads to a skewness exceeding $2$. The opposite however is not possible, if the constraint $\S\le 2$ is satisfied, the
randomness parameter never goes above $1$.              

A randomness parameter larger than one can also be obtained in a case where there is only one cycle but with an extra state that does not belong to the cycle \cite{moff10a,saha12}. Such 
an off-pathway state may be related to an inhibitor. This situation is already contained in the mathematical model defined in eqs (\ref{reaction1}) and (\ref{reaction2}): by setting the rates $k_{34}=k_{43}=0$
and identifying the states $\textrm{E}\mathrel{\hat=} 1$, $\textrm{ES}\mathrel{\hat=} 2$, $\textrm{EI}\mathrel{\hat=} 3$, and $\textrm{EP}\mathrel{\hat=} 4$ we obtain the enzymatic scheme in Figure \ref{fig1.5b}.
Our numerics demonstrate that also for this case with an off-pathway state due to an inhibitor higher order moments are more effective in revealing that the network is not just a single cycle. 
It remains as an open question whether this property is true for arbitrary networks that do not consist of a single cycle.


In conclusion, we have obtained lower and upper bounds on the skewness and kurtosis associated with the cycle completion time.
These bounds can lead to information about the enzymatic scheme that is not contained in the randomness parameter.
By evaluating skewness and kurtosis for experimental data further information on the number of intermediate
states and on whether the network of states is more complex than a cycle can now be obtained. While for the former information excellent statistics might be needed to evaluate 
higher order moments so that the upper bounds on the differences $\D$ can be verified, for the latter even rough
measures of higher order moments may be enough: it is possible to find parameter regimes where the randomness parameter 
is close to 1 but the kurtosis and skewness are way above their maximum values for unicyclic schemes. 

On the theoretical side, it is intriguing to ask whether general expressions for the skewness and kurtosis 
in terms of the substrate concentration, as the expression for the randomness parameter 
derived in \cite{moff10}, exists and what kind of information about the enzymatic scheme such expressions 
would reveal. Second, it would be worthwhile to consider higher order moments associated with the number of generated product molecules 
for thermodynamic consistent models, i.e., models without
irreversible transitions. In this case stronger lower bounds
on the number of intermediate states depending on the chemical potential difference driving the reaction exist \cite{bara14,bara15}.




\begin{thebibliography}{10}
\bibitem{foot} The Fano factor, which is related to the randomness parameter \cite{moff14}, can be larger
than one in reversible models with a small chemical potential difference that drives the reaction 
\cite{fish01,kolo03,bara14,bara15}.
\end{thebibliography}

\clearpage

\onecolumngrid
\section*{Supplemental Material: Skewness and Kurtosis in Statistical Kinetics}

\section{Relation between cycle completion time distribution and the probability of product generation}

The relations derived in this section are valid for both the unicyclic scheme in eq (1) and for the multicyclic 
scheme from eqs (18) and (19). The basic assumption that must hold is that the step leading to product formation is
irreversible and common to all cycles. We denote the probability of producing $M$ products until time $t$ by $P(M,t)$. 
The cycle completion time probability $\psi(t)$ is related to the survival probability    
\begin{equation}
P_s(t)\equiv 1-\int_0^t \psi(t')dt',
\tag{S1}
\end{equation}
which is the probability of not completing a cycle until time $t$.
These three probabilities are connected with the expression
\begin{equation}
P(1,t)=  \int_0^t \psi(t')P_s(t-t')dt',
\tag{S2}
\label{eq1}
\end{equation}
which simply means that the probability of completing one cycle up to time $t$ is the probability of completing a cycle at
time $t'$ multiplied by the probability of surviving for $t-t'$ integrated over $t'\le t$. For $M>1$ we have the recursion relation 
\begin{equation} 
P(M,t)=  \int_0^t \psi(t')P(M-1,t-t')dt'.
\tag{S3}
\label{eqM}
\end{equation}
With the generating functions $\tilde{\psi}(s)$, defined in eq (7) in the main text, and $\tilde{P}(M,s)\equiv\int_0^\infty P(M,t)\textrm{e}^{-st}$, eqs (\ref{eq1}) and (\ref{eqM}) lead to  
\begin{equation}
\tilde{P}(M,s)= [\tilde{\psi}(s)]^M\frac{1-\tilde{\psi}(s)}{s}.
\tag{S4}
\label{eqms}
\end{equation}
From (\ref{eqms}), with the definition
\begin{equation}
\tilde{P}(z,s)\equiv \sum_{M=0}^{\infty}\tilde{P}(M,s)\textrm{e}^{zM},
\tag{S5}
\end{equation}
we obtain the relation 
\begin{equation}
\tilde{P}(z,s)= \frac{1-\tilde{\psi}(s)}{s[1-\textrm{e}^z\tilde{\psi}(s)]}.
\label{ppsi}
\tag{S6}
\end{equation}
In the context of enzyme kinetics this relation has been derived in \cite{schn95}.

The master equation can be written in the form \cite{koza99,lebo99}  
\begin{equation}
\frac{d}{d\tau}\mathbf{P}(z)= \mathbf{L}(z)\mathbf{P}(z),
\tag{S7}
\label{mastereq2}
\end{equation}
$\mathbf{P}(z)$ is a vector with components $P_i(z,t)$, with $i=1,2,\ldots,N$. These components are defined as 
\begin{equation}
P_i(z,t)\equiv \sum_{M=0}^{\infty}P_i(M,t)\textrm{e}^{zM},
\tag{S8}
\end{equation}
where $P_i(M,t)$ is the probability that $M$ cycles have been completed until time $t$ and the enzyme
is in the intermediate state $i$. This probability is related to $P(M,t)$ through the expression $P(M,t)=\sum_{i=1}^{N}P_i(M,t)$. 
With a Laplace transform in time, for which $t\to s$, eq (\ref{mastereq2}) becomes 
\begin{equation}
\mathbf{\tilde{P}}(z,s)= [s\mathbf{I}-\mathbf{L}(z)]^{-1}\mathbf{P}(0),
\label{pzndef}
\tag{S9}
\end{equation}
where the components of the initial condition vector $\mathbf{P}(0)$ are $P_i(0)= \delta_{1,i}$. Eqs (\ref{ppsi}), where $\tilde{P}(z,s)$ is the sum of the components of the vector $\mathbf{\tilde{P}}(z,s)$, and (\ref{pzndef})  
 are used in the derivation of expression (10) in the main text, which has been obtained in \cite{chem08}.

For the multicyclic model in Eqs (18) and (19) in the main text, the modified generator in eq. (\ref{mastereq2}) reads
\begin{equation}
\mathbf{L}(z)\equiv\left(
\begin{array}{cccc}
-r_1 & k_{21} & k_{31} & k_{41}\textrm{e}^z \\
 k_{12}         & -r_2 & 0 & k_{42} \\ 
 k_{13}          & 0 & -r_3 & k_{43} \\
 0          & k_{24} & k_{34} & -r_4
\end{array}
\right),  
\label{matrix1}
\tag{S10}
\end{equation}
where $r_1\equiv k_{12}+k_{13}$, $r_2\equiv k_{21}+k_{24}$, $r_3\equiv k_{31}+k_{34}$, and $r_4\equiv k_{41}+k_{42}+k_{43}$. 
We can then obtain $\tilde{\psi}(s)$ by solving eq (\ref{ppsi}) after calculating $\tilde{P}(z,s)$ from eqs  (\ref{pzndef}) and (\ref{matrix1}).
With  $\tilde{\psi}(s)$ we obtain skewness and kurtosis from eqs (3), (5), and (6) in the main text. The general expressions
are quite long.

\section{Proof for $N=2$ and Numerical Evidence}

For the case $N=2$ the expressions for $\S$ and $\K$ are simple and the bounds (12) and (13) in the main text can be proven explictly as follows.
These expressions are obtained as explained in the main text after eq (10). The skewness reads
\begin{equation}
\S= 2 (A^3-3k_1^+A)/B^{3/2}
\tag{S11}
\end{equation}
where $A\equiv 1+k_1^++k_1^-$ and $B\equiv A^2-2k_1^+$. Taking the derivative with respect to $k_1^-$ we obtain
\begin{equation}
\frac{\partial{\S}}{\partial k_1^-}= 12 (k_1^+)^2/B^{5/2}\ge 0.
\tag{S12}
\end{equation}
Hence, $\S$ is a monotonically  increasing function of $k_1^-$. The  derivative with respect to $k_1^+$ gives
\begin{equation}
\frac{\partial{\S}}{\partial k_1^+}= - \frac{6(k_1^+-k_1^--1)}{B^{5/2}}.
\tag{S13}
\end{equation}
From these two equations it follows that $\S$ is minimized at $k_1^+=1$ and $k_1^-=0$, where it becomes $\S=\sqrt{2}$. The skewness is maximized
for $k_1^+$ finite and $k_1^-\to \infty$, where it becomes $\S=2$. These results are in agreement with the bounds in eq (12) in the main text.
The kurtosis reads
\begin{equation}
\K= \frac{3[8(k_1^+)^2-12k_1^+A^2+3A^4]}{B^{2}}.
\tag{S14}
\end{equation}
The derivatives are given by
\begin{equation}
\frac{\partial{\K}}{\partial k_1^-}= \frac{48(k_1^+)^2A}{C^{3}}\ge 0
\tag{S15}
\end{equation}
and
\begin{equation}
\frac{\partial{\K}}{\partial k_1^+}= -\frac{24k_1^+(k_1^+-k_1^--1)^2}{C^{3}},
\tag{S16}
\end{equation}
where $C\equiv 1+(k_1^+)^2+(k_1^-)^2+2k_1^-(1+k_1^+)$. Similar to the skewness, the kurtosis is minimized at $k_1^+=1$ and $k_1^-=0$, where it becomes $\K=6$, and maximized for
$k_1^+$ finite and $k_1^-\to \infty$, where it becomes $\K=9$. These results are in agreement with the bounds in eq (13) in the main text.

We cannot perform the same explicit proof for larger values of $N$ as both functions become more complicated as $N$ increases. We did evaluate analytically the derivatives of $\S$ and $\K$ with respect
to one of the forward rates at the point where all forward rates are 1 and the backward rates are zero. In all cases this derivative is 0, in agreement with $\S$ and $\K$ being minimized at this point,
where they reach the lower bounds in eqs (12) and (13) in the main text.

\begin{figure}
\subfigure[]{\includegraphics[width=72mm]{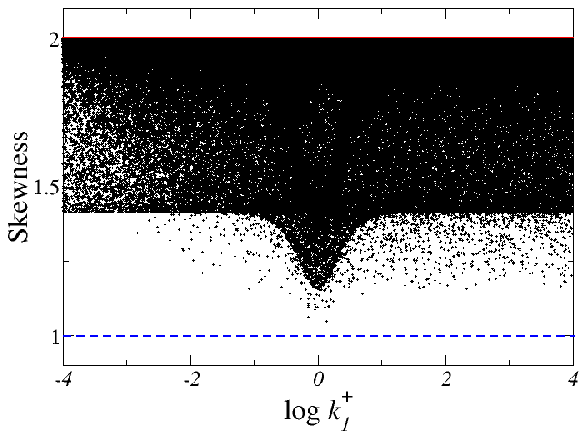}\label{figA1a}}\hfill
\subfigure[]{\includegraphics[width=72mm]{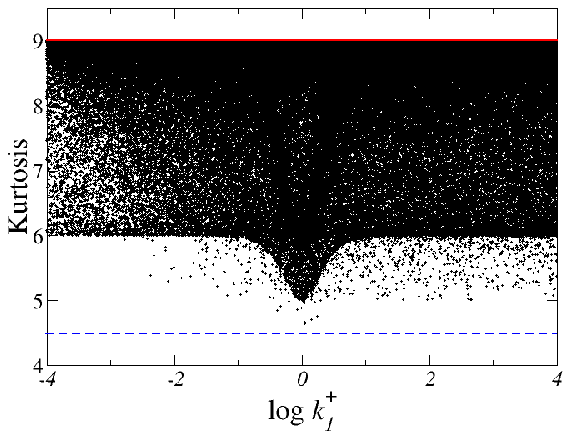}\label{figA1b}}\hfill
\subfigure[]{\includegraphics[width=72mm]{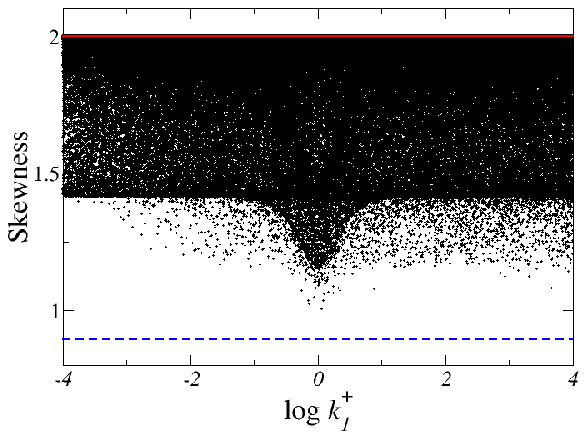}\label{figA1c}}\hfill
\subfigure[]{\includegraphics[width=72mm]{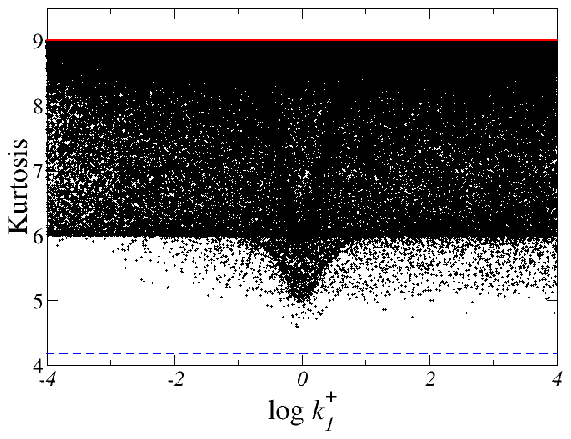}\label{figA1d}}\hfill
\subfigure[]{\includegraphics[width=72mm]{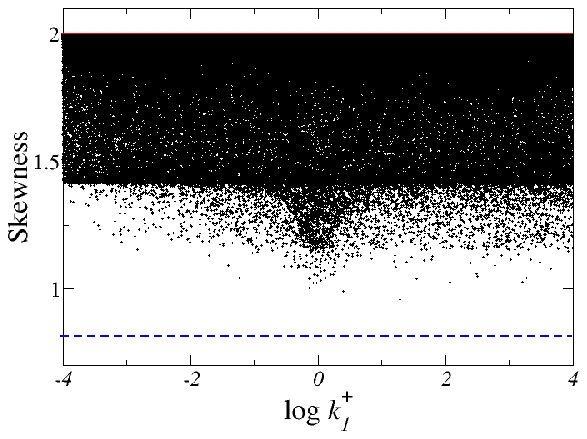}\label{figA1e}}\hfill
\subfigure[]{\includegraphics[width=72mm]{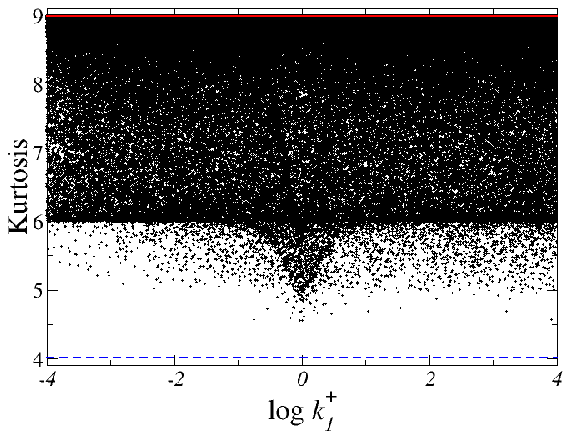}\label{figA1f}}
\vspace{-2mm}
\caption{Scatter plots illustrating inequalities (14) and (15) in the main text. The solid red line represents the upper bounds and the dotted blue line the lower bounds in these equations. 
$N=4$ for Figures \ref{figA1a} and \ref{figA1b},  $N=5$ for Figures \ref{figA1c} and \ref{figA1d}, and $N=6$ for Figures \ref{figA1e} and \ref{figA1f}. For all figures $k_N^+=1$ and
the other transition rates are taken as $10^x$, with $x$ a random number between $-4$ and $4$. Each scatter plot has $10^6$ points. The minima close to $k_1^+=1$ comes from
the fact that skewness and kurtosis are minimized when all forward rates are $1$ and the backward rates are $0$.
}
\label{figA1} 
\end{figure}

The definite evidence for the bounds come from numerical minimization and maximization of $\S$ and $\K$ from $L=2$ up to $L=8$. As an independent check we have also
evaluated these functions at randomly chosen transition rates. Some of the results obtained with this second procedure are shown in Figure \ref{figA1}.

\end{document}